\documentclass
[pra,superscriptaddress,showpacs,twocolumn,aps,floatfix,
  10pt]{revtex4-1}%
\usepackage{hyperref}
\usepackage[pdftex]{graphicx}
\usepackage{bm}
\usepackage{etoolbox}
\usepackage[usenames,dvipsnames]{color}
\usepackage{amsmath}
\usepackage{amsfonts}
\usepackage{amssymb}%
\newcommand{\bg}{ \begin{gather} }
\newcommand{\eg}{\end{gather}}
\newcommand{\be}{ \begin{equation} }
\newcommand{\ee}{\end{equation}}
\newcommand{\bea}{ \begin{eqnarray} }
\newcommand{\eea}{\end{eqnarray}}

\begin{document}

\title{Resonant supercollisions and electron-phonon heat transfer in graphene}

\author{K.\,S.~Tikhonov}
\affiliation{Institut f\"ur Nanotechnologie,  Karlsruhe Institute of Technology,
76021 Karlsruhe, Germany}
\affiliation{L.\,D.~Landau Institute for Theoretical Physics, 119334 Moscow, Russia}
\affiliation{National University of Science and Technology "MISiS", 119049 Moscow, Russia}
\author{I. V.~Gornyi}
\affiliation{Institut f\"ur Nanotechnologie,  Karlsruhe Institute of Technology,
76021 Karlsruhe, Germany}
\affiliation{A. F.~Ioffe Physico-Technical Institute,
194021 St.~Petersburg, Russia}
\affiliation{\mbox{Institut f\"ur Theorie der kondensierten Materie,  Karlsruhe Institute of
Technology, 76128 Karlsruhe, Germany}}
\affiliation{L.\,D.~Landau Institute for Theoretical Physics, 119334 Moscow, Russia}

\author{ V. Yu.~Kachorovskii}
\affiliation{A. F.~Ioffe Physico-Technical Institute,
194021 St.~Petersburg, Russia}
\affiliation{L.\,D.~Landau Institute for Theoretical Physics, 119334 Moscow, Russia}
\affiliation{Institut f\"ur Nanotechnologie,  Karlsruhe Institute of Technology,
76021 Karlsruhe, Germany}

\author{A. D.~Mirlin}
\affiliation{Institut f\"ur Nanotechnologie,  Karlsruhe Institute of Technology,
76021 Karlsruhe, Germany}
\affiliation{\mbox{Institut f\"ur Theorie der kondensierten Materie,  Karlsruhe Institute of
Technology, 76128 Karlsruhe, Germany}}
\affiliation{Petersburg Nuclear Physics Institute, 188300, St.Petersburg, Russia}
\affiliation{L.\,D.~Landau Institute for Theoretical Physics, 119334 Moscow, Russia}

\begin{abstract}
We study effects of strong impurities on the heat transfer in a
coupled electron-phonon system in disordered graphene. A detailed analysis of the electron-phonon heat exchange assisted by such an impurity through the ``resonant supercollision'' mechanism is presented. We further explore the local modification of heat transfer in a weakly disordered graphene due to a resonant scatterer and determine spatial profiles of the phonon and electron temperature around the scatterer under electrical driving.
Our results  are consistent with recent experimental findings on imaging resonant dissipation from individual atomic defects.
\end{abstract}
\maketitle

\section{Introduction}
\label{s1}

Dissipation of energy in electron transport in nanostructures is of fundamental interest and of importance for applications. At low temperatures, the electric resistance is usually governed by elastic scattering off impurities. This resistance determines, in particular, the amount of Joule heat (for a given applied voltage or current). However, the heat dissipation requires an energy transfer from the electronic system to the ``environment''---usually, to phonons. Thus, understanding the character of heat dissipation is a complex problem which requires an analysis of the electron-phonon scattering and, more generally, of the heat transfer in a system of electrons and phonons.  Remarkably, the heat dissipation (i.e., the delivery of the energy gained by electrons in the electric field to phonons) may be even spatially separated from the region which dominates the resistance, as is the case for a ballistic point contact \cite{rokni1995joule}. Recent work \cite{halbertal2016nanoscale} has developed a highly sensitive experimental technique of thermal nanoimaging which utilises a superconducting  quantum interference device (SQUID) located on a tip. This technique allows one to obtain a spatial temperature distribution with a resolution of order of micro-Kelvin in temperature and of
order of nanometer in space.

The character of dissipation is of special interest in the case of graphene which represents an ultimate two-dimensional (2D) material. It was shown that at sufficiently high temperatures, the dominant electron-phonon relaxation processes are ``supercollisions'' assisted by impurities~\cite{song2012disorder,song2015energy}, as has been also confirmed experimentally~\cite{betz2012supercollision}. Related studies of the electron-phonon cooling rates were reported in Refs.~\onlinecite{mckitterick2016electron,fong2013measurement}.

A very recent experiment has reported remarkable results of thermal imaging of dissipation on graphene~\cite{2017arXiv171001486H}. Specifically, the authors of Ref.~\cite{2017arXiv171001486H} observed dissipation ``hot spots'' and provided strong evidence that they are associated with individual resonant impurities. It is indeed known that in graphene strong impurities induce resonances near the Dirac point and may crucially affect transport properties~\cite{ostrovsky2006electron,pereira2006disorder,basko2008resonant,titov2010charge}. The technique of Ref.~\cite{2017arXiv171001486H} has permitted to  observe ``dissipation rings'' in the thermal image which correspond to positions of the tip \cite{halbertal2016nanoscale} at which the individual defect is at resonance (at given values of the back-gate voltage and the tip voltage).

The goal of this work is to study the effect of resonant impurities on the heat transfer in a coupled electron-phonon system in graphene.
In general, the effect of a strong impurity on the energy dissipation around it may be twofold.
First, the impurity modifies locally the electron-phonon collision rate, leading to ``resonant supercollisions'' that we explore in Sec.~\ref{s2}.
Remarkably, this effect is drastically enhanced  in graphene  due to the relativistic character of its spectrum,  which leads to a strong singularity of the impurity-scattering waves:  $\Psi_{\rm scat} \propto  1 /r$ as compared to $ \Psi_{\rm scat} \propto \ln r$
for conventional 2D semiconductors with a parabolic spectrum.
As a result, the intensity of energy exchange between electrons and phonons is enhanced, the excessive momenta being
transferred  to the impurity.
The effect shows up already within the Born approximation (with respect to the impurity potential),
leading to ``weak'' supercollisions \cite{song2012disorder}.
The corresponding phonon matrix element  $M(\mathbf q) \propto \int d^2r  \langle \Psi_{\rm scat}|  \exp(i\mathbf q\mathbf r)|\mathbf k \rangle $
slowly decays with $q$  at $q\gg k_F$: $M(\mathbf q) \propto 1/q$
and the electron-phonon heat flux scales as $T^3 \delta^2 $  (here   $|\mathbf k\rangle$ is  the plane wave and $\delta  \ll 1$ is the scattering phase).
In this paper, we demonstrate that beyond the Born approximation, the matrix element decays much slower:
$M(\mathbf q) \propto \int d^2r  \langle \Psi_{\rm scat}|  \exp(i\mathbf q\mathbf r)| \Psi_{\rm scat}\rangle\propto \ln (1/qR )$ (here $R$ is the impurity size) and the heat flux dramatically increases with increasing temperature, scaling as $T^5 \sin^4 \delta$ (up to logarithmic factors).
For weak impurities, $\delta \ll 1$, this effect overcomes the effect of Born-approximation supercollisions \cite{song2012disorder}
at sufficiently large temperatures. For strong impurities with $\delta \sim 1$ the resonant contribution dominates the impurity-mediated heat flux at all temperatures.

The second effect of an individual impurity on the energy dissipation is a local modification of the electric field and current profiles and thus of the associated Joule heat. Such a modification of field by a scatterer is associated in the literature with the notion of ``Landauer residual-resistivity dipoles'' \cite{landauer1957spatial,landauer1988spatial,chu88,zwerger1991exact,sorbello1998landauer}. Imaging techniques permit a direct observation of such dipoles by measurement of the spatial distribution of current and voltage on nanoscale~\cite{homoth2009electronic,willke2015spatial}. In Sec.~\ref{s3} we formulate a heat-transfer model that takes into account both kinds of effects induced by an impurity and determine a local profile of electronic and phonon temperatures around a scatterer under electrical driving. As we show in Appendix \ref{App:A}, the effect of additional Joule heating due to ``Landauer dipoles'' is relatively small in 2D systems, so that the modification of heating near the impurity is predominantly due to the effect of the impurity on the electron-phonon scattering.
The estimates of a characteristic magnitude of the effect for realistic experimental parameters, as well as a comparison to the experiment of Ref.~\cite{2017arXiv171001486H} is presented in Sec.~\ref{s3B}. In Sec.~\ref{s4}, we summarize our results.
Throughout the paper, we set $\hbar=k_B=1$ in some intermediate formulas and restore these constants in final expressions.

\section{Supercollisions on resonant impurities}
\label{s2}

\subsection{Impurities in graphene}
\label{s2.1}

We start with the $4\times 4$ Dirac Hamiltonian for graphene
\begin{equation}
\hat{\mathcal{H}}=v \hat{\tau}_3 \hat{\bm{\sigma}} \mathbf{k},
\end{equation}
where $v$ is the Dirac velocity,
$\hat{\tau}_3$ is the Pauli matrix acting in the valley space ($K,\, K'$)
and $\hat{\bm{\sigma}}$ is the vector of Pauli matrices in the sublattice space ($A,\, B$).
Electronic states are given by vectors of amplitudes
\begin{equation}
\psi=(\psi_{AK},\psi_{BK},\psi_{BK^{\prime}},\psi_{AK^{\prime}})^{T}.%
\end{equation}
Scattering of an electron with energy $\epsilon=v|\mathbf{k}|$
(wavevector $\mathbf{k}$ of the incident wave is counted from the Dirac point $\alpha=K,\, K'$) on a single
impurity
centered at position $\mathbf{r}=0$ is described by the wavefunction
\begin{equation}
\psi_{\mathbf{k}}(\mathbf{r})=\left[e^{i\mathbf{k}\mathbf{r}}+\hat{t}(\epsilon,r)\right]  \left\vert\mathbf{k}\alpha\right\rangle.
 \label{psi}%
\end{equation}
Here $\hat{t}(\epsilon,r)$ is the transfer matrix
and the spinors $\left\vert\mathbf{k}\alpha\right\rangle$ depend on the direction $\phi_{k}$ of the
electron momentum:
\begin{eqnarray}
\left\vert\mathbf{k}K\right\rangle
&=&\frac{1}{\sqrt{2}}\left(1,e^{i\phi_{k}},0,0\right)^{T} \label{psik},\\
\left\vert\mathbf{k}K^\prime\right\rangle
&=&\frac{1}{\sqrt{2}}
\left(  0,0,1,-e^{i\phi_{k}}\right)^{T}. \label{psikp}%
\end{eqnarray}

We consider the two types of impurity potential with the spatial extension
smaller than the Fermi wavelength: ``atomically sharp'' (short-range)
and ``atomically smooth'' (long-range on the scale of the lattice constant).
The transfer matrix takes the form
\begin{equation}
\hat{t}(\epsilon,r)=\frac{\hat{G}(\epsilon,r)\hat{U}}{1-\hat{G}(\epsilon,0)\hat{U}}
\label{tr}%
\end{equation}
and describes the $s$-wave scattering off impurity.
For a short-range (long-range) impurity centered
at the site of sublattice $A$, the potential has the following matrix structure:
\begin{equation}
\hat{U}_\text{sr}=2U_{0}\hat{\Lambda}_\text{sr},\quad \hat{U}_\text{lr}=U_{0}\hat{\Lambda}_\text{lr}%
\end{equation}
with the amplitude $U_0$ and
\begin{equation}
\hat{\Lambda}_\text{sr}=\left(
\begin{array}
[c]{cccc}%
1/2 & 0 & 0 & 1/2\\
0 & 0 & 0 & 0\\
0 & 0 & 0 & 0\\
1/2 & 0 & 0 & 1/2
\end{array}
\right), \quad \hat{\Lambda}_\text{lr}=\left(
\begin{array}
[c]{cccc}%
1 & 0 & 0 & 0\\
0 & 1 & 0 & 0\\
0 & 0 & 1 & 0\\
0 & 0 & 0 & 1
\end{array}
\right). \label{Lambda}%
\end{equation}
The Green function $\hat{G}$ at energy $\epsilon$ reads
\begin{align}
\hat{G}(\epsilon,r)  &
 =\int\frac{d^{2}k}{(2\pi)^{2}}
 \frac{\epsilon +v\mathbf{k\hat{\bm{\sigma}}}\hat{\tau}_{3}}{\epsilon^{2}-v^{2}k^{2}+i0}
 e^{i\mathbf{k}\mathbf{r}}\\
&
=
\frac{1}{4iv^{2}}
\left[  \epsilon+v\hat{\tau}_{3}{\hat{\bm{\sigma}}}(-i\partial_{r})\right]
H_{0}^{(1)}(\epsilon r/v),
\end{align}
where $H_0^{(1)}(z)$ is the Hankel function of the first kind.
For finding the transfer matrix, we make use of the small-$r$
expansion:
\begin{equation}
\hat{G}(\epsilon,r\rightarrow0)\simeq
-\frac{1}{2\pi v}
\left[  i\hat{\tau}_{3}
\frac{\hat{\bm{\sigma}}\mathbf{r}}{r^{2}}
+\frac{\epsilon}{v}\left(  \ln\frac{v}{\epsilon r}+\frac{i\pi}{2}\right)  \right].
\label{G}
\end{equation}
The limit $\hat{G}\left(\epsilon,r\rightarrow 0\right)$ in the denominator of Eq.~(\ref{tr}) is taken
after the integration over the spatial region where the impurity potential is nonzero.
For an isotropic impurity of small radius, the first term in Eq. (\ref{G}) does not contribute to Eq.~(\ref{tr})
because of the angular integration,
\begin{equation}
\hat{G}\left(\epsilon,0\right) \rightarrow
-\frac{\epsilon}{2\pi v^{2}}\left(
\ln\frac{v}{\epsilon R}+\frac{i\pi}{2}\right), \label{G0}%
\end{equation}
where $R$ is the ultraviolet scale (the radius of the scattering potential or the lattice constant,
whichever is larger).
Performing matrix operations, we obtain
\begin{equation}
\hat{t}(\epsilon,r)=\frac{4v^{2}\sin\delta e^{-i\delta}}{\epsilon}
\hat{G}(\epsilon,r)\hat{\Lambda},
\label{tm}
\end{equation}
where $\hat{\Lambda}$ equals either
$\hat{\Lambda}_\text{lr}$ or $\hat{\Lambda}_\text{sr}$
and the scattering phase $\delta$ is governed by the strength of the impurity,
\begin{equation}
\cot\delta=\frac{v}{\epsilon L}+\frac{2}{\pi}\ln\frac{v}{|\epsilon| R},
\end{equation}
with the scattering length $L$ given by
\begin{equation}
L_\text{lr}=\frac{U_{0}}{4v},\quad L_\text{sr}=\frac{U_{0}}{2v} \label{Ls}%
\end{equation}
for the long-range or short-range case, respectively.
For a strong impurity, $L \gg R$,
the transfer matrix (\ref{tm})
acquires a resonant energy dependence
\begin{equation}
\sin^2\delta \simeq \frac{\Gamma^2}{(\epsilon-\epsilon_\text{res})^2+\Gamma^2}
\end{equation}
with the resonant energy
\begin{equation}
\epsilon_\text{res}\simeq -\frac{\pi v}{2 L \ln(L/R)}
\end{equation}
and the width
\begin{equation}
\Gamma\simeq \frac{\pi^2 v}{4 L \ln^2(L/R)}\ll |\epsilon_\text{res}|.
\end{equation}
In what follows, we refer to such impurities as resonant ones; note that the stronger the impurity,
the closer the resonant energy to the Dirac point ($\epsilon_\text{res}\to 0$ for $U_0\to \infty$)
and the sharper the resonance.
Below, we will analyze the electron-phonon interaction in the presence of
impurities and show that scattering off a resonant impurity may strongly
enhance the heat exchange between electrons and phonons.

\subsection{Impurity-assisted electron-phonon scattering}
\label{s2.2}

Let us now consider the matrix element of electron-phonon scattering in graphene in the presence
of an isolated impurity.
Since, by assumption, the impurity potential
is strong, it can not be treated perturbatively. Instead,
we calculate phonon-induced scattering between
exact impurity-scattering states (\ref{psi}).
Assuming two-dimensional phonons with the phonon wavevector $\mathbf{q}$ in the graphene plane,
the matrix element of $\exp(i\mathbf{q}\mathbf{r})$ reads:
\begin{equation}
M^{\alpha\beta}_{\mathbf{k}\mathbf{k}^{\prime}}\!(\mathbf{q})\!
=\!\left\langle \mathbf{k}\alpha\right\vert\!
\left[e^{-i\mathbf{k}\mathbf{r}}\!+\hat{t}^{\dagger}\left(\epsilon,r\right)  \right]\!  e^{i\mathbf{q}\mathbf{r}}
\!\left[e^{i\mathbf{k}^\prime\mathbf{r}}\!+\hat{t}\left(\epsilon^\prime\!,r\right)\right]\!
\left\vert \mathbf{k}^{\prime}\beta\right\rangle,
\label{Mdef}%
\end{equation}
where $\alpha,\beta$ denote the valleys.

We will focus on the case of large phonon momenta $q\gg k,k^{\prime}$,
when the spatial structure of the electronic wavefunctions is irrelevant.
The effect of impurity (a supercollision \cite{song2012disorder})
can be represented as a sum of the two terms:
\begin{eqnarray}
M^{\left(  1\right)  }
&=&\frac{4v^{2}\sin\delta}{\epsilon}
\label{M1def}\\
&\times&
\left\langle \mathbf{k}{\alpha}\right\vert
e^{i\mathbf{q}\mathbf{r}}
\left[e^{i\delta}\hat{\Lambda}\hat{G}^{\dagger}
\left(\epsilon,r\right)+  e^{-i\delta}\hat{G}\left(\epsilon^\prime,r\right)
\hat{\Lambda} \right]
\left\vert \mathbf{k}^{\prime}{\beta}\right\rangle
\notag
\end{eqnarray}
and%
\begin{equation}
M^{\left(  2\right)  }
=
\frac{16v^{4}\sin^{2}\delta}{\epsilon^{2}}
\left\langle \mathbf{k}{\alpha}\right\vert e^{i\mathbf{q}\mathbf{r}}
\hat{\Lambda}\hat{G}^{\dagger}\left(\epsilon,r\right)
\hat{G}\left(\epsilon^\prime,r\right)
\hat{\Lambda} \left\vert \mathbf{k}^{\prime}{\beta}\right\rangle.
\label{M2def}
\end{equation}
As we will see below, only the short-distance asymptotics of the Green function should be kept,
as the supercollision matrix element
at large $q$ is dominated by the most singular (at $r\rightarrow0$)
terms in $G(\epsilon,r)$.
In what follows, we assume $T\ll E_{F}$, so that
$\left\vert k\right\vert
\approx\left\vert k^{\prime}\right\vert $.
For $M^{\left(  1\right)}$ we then
obtain%
\begin{eqnarray}
M^{\left(  1\right)}&=&
\frac{4v^{2}\sin\delta}{\epsilon}
\label{M1appr}%
\\
&\times&
\left\langle
\mathbf{k}{\alpha}\right\vert \cos\delta\,
\left[  \hat{\Lambda},\hat{G}_\mathbf{q}\right]
+i\sin\delta\, \left\{  \hat{\Lambda},\hat{G}_\mathbf{q}\right\}
\left\vert \mathbf{k}^{\prime}\beta\right\rangle,
\notag
\end{eqnarray}
in terms of the commutator and anticommutator of matrix $\hat{\Lambda}$
with the Fourier-transformed Green function $\hat{G}_\mathbf{q}$.
Using the asymptotics of $\hat{G}_\mathbf{q}$ at large $q$%
\begin{equation}
\hat{G}_\mathbf{q}\simeq -\frac{\left( \hat{\bm{\sigma}} \mathbf{ q}\right)  }{vq^{2}}\hat{\tau}_{3},
\label{Gappr}%
\end{equation}
we obtain
\begin{align}
M^{\left(  1\right)  }  &
\simeq
-\frac{4v\sin\delta\cos\delta}{\epsilon q}
\left\langle \mathbf{k}{\alpha}\right\vert
\left[  \hat{\Lambda},\left(\hat{\bm{\sigma}}\hat{\mathbf{q}}\right)  \hat{\tau}_{3}\right]
\left\vert \mathbf{k}^{\prime}{\beta}\right\rangle
\label{M1}\\
&
-\frac{4iv\sin^{2}\delta}{\epsilon q}
\left\langle \mathbf{k}{\alpha}\right\vert
\left\{  \hat{\Lambda},\left(  \hat{\bm{\sigma}}\hat{\mathbf{q}}\right)
\hat{\tau}_{3}\right\}
\left\vert \mathbf{k}^{\prime}{\beta}\right\rangle,\nonumber
\end{align}
where $\mathbf{\hat{q}}=\mathbf{q}/q $.
For $M^{\left(2\right)}$, we notice that the most singular (at $r\rightarrow0$)
term in the product $\hat{G}^{\dagger}\hat{G}$ reads as follows:
\begin{equation}
\hat{G}^{\dagger}\left(\epsilon,r\right)  \hat{G}\left(\epsilon^\prime,r\right)
\approx\frac{1}{4\pi^{2}v^{2}r^{2}}.
\label{GG}%
\end{equation}
As a result, using $\hat{\Lambda}^{2}=\hat{\Lambda}$, we obtain
\begin{equation}
M^{\left(  2\right)  }
\simeq
\frac{8v^{2}\sin^{2}\delta}{\pi\epsilon^{2}}
\ln\frac{1}{qR}\left\langle \mathbf{k}{\alpha}\right\vert
\hat{\Lambda}\left\vert \mathbf{k}^{\prime}{\beta}\right\rangle .
\label{M2}
\end{equation}

In the Born approximation (to the lowest order in $\delta\ll 1$),
only the first line of Eq.~(\ref{M1}) is present.
This contribution was calculated in Ref. \cite{song2012disorder}. Remarkably, the
contribution of Eq.~(\ref{M2}), absent in the Born approximation, decreases
with $q$ much slower. As a result, it dominates the electron-phonon heat exchange
at sufficiently high temperatures
even for weak impurities (the second line of Eq.~(\ref{M1}) is always small).
Indeed, the ratio of the respective contributions to the matrix element at
thermal phonon wavevectors $q_{T}=T/s$
is given by (omitting coefficients of order unity)
\begin{equation}
\frac{M^{\left(  2\right)  }}{M^{\left(  1\right)  }}
\sim \tan\delta\ \frac{T}{T_{BG}}
\ln\frac{1}{q_{T}R},
\label{Mratio}
\end{equation}
where
\begin{equation}
T_\text{BG}=\hbar sk_{F}/k_B%
\simeq 27\text{K} \sqrt{n/10^{12}\text{cm}^{-2}}
\label{TBG-def}
\end{equation}
is the Bloch-Gr\"uneisen temperature ($s$ is the sound
velocity, $k_F$ the Fermi momentum, and $n$ the electron concentration).

The condition $q\gg k,k^\prime$ for typical phonons is realized at sufficiently high temperatures $T\gg T_\text{BG}$.
When this condition is not fulfilled, the supercollision matrix elements can be estimated as
(dropping the numerical coefficients)
\begin{eqnarray}
M^{\left(1\right)}(q\ll k,k^\prime) &\sim& \frac{\sin \delta \cos\delta}{k_F^2},
\label{M1-smallq}
\\
M^{\left(2\right)}(q\ll k,k^\prime) &\sim&  \frac{\sin^2 \delta}{k_F^2} \ln\frac{1}{k_FR},
\label{M2-smallq}
\end{eqnarray}
and hence their ratio at $T\ll T_\text{BG}$ is given by Eq.~(\ref{Mratio}) with $T\sim T_\text{BG}$.
For a strong impurity $\tan\delta\gtrsim 1$.  Since the argument of logarithm in Eq.~(\ref{Mratio})
is always large,
for resonant impurities one has $M^{\left(2\right)}>M^{\left(1\right)}$
in the whole temperature range.

\subsection{Heat flux between electrons and phonons}
\label{s2.3}

Let us now evaluate the impurity-assisted heat flux $J$ from electrons to phonons.
The Fermi golden rule yields (cf. Ref.~\cite{song2012disorder})
\begin{align}
J&=\frac{2\pi}{\hbar}\,n_\text{imp}\, \nu^{2}_F\,g^{2}
\label{Jdef}\\
&\times2\sum_{\alpha\beta}\int\frac{d^{2}q}{\left(  2\pi\right)^{2}}
\omega_{q}^{3}\left(  N_{\omega_{q}}^\text{e}-N_{\omega_{q}}^\text{ph}\right)
\left\langle
\left\vert M^{\alpha\beta}_{\mathbf{k}\mathbf{k}^{\prime}}\left(\mathbf{q}\right) \right\vert^{2}
\right\rangle_\text{FS},\nonumber
\end{align}
where $n_\text{imp}$ is the impurity concentration, $g=D/\sqrt{2\rho s^{2}}$ the electron-phonon coupling constant  (with $D$ being
the deformation-potential constant and $\rho$ the graphene mass density), $\nu_F=k_F/(2\pi\hbar v_F)$ the electronic density of states
per spin per valley at the Fermi level, $\omega_q=s q$ the phonon dispersion, and $\left\vert k\right\vert =\left\vert k^{\prime}\right\vert =k_F$. Further,
$\left\langle \ldots\right\rangle_\text{FS} $ stands for the Fermi-surface averaging over angles of
$\mathbf{k},\,\mathbf{k}^{\prime}$, and $N_{\omega}^\text{e,ph}$ are the Bose distribution functions with
electron and phonon temperatures, $T_\text{e}$ and $T_\text{ph}$, respectively.

Performing the integration over $q$ with Eqs.~(\ref{M1}) and (\ref{M2}) for the matrix element,
we arrive at
\begin{equation}
J=I\left(T_\text{e}\right)-I\left( T_\text{ph}\right),  \label{Idef}
\end{equation}
where
\begin{equation}
I\left(T\right)=I_0(T)+I_\text{Born}(T)+I_\text{res}(T).
\end{equation}
In this expression, $I_0(T)$ is the contribution that does not involve
the impurity scattering, the term $I_\text{Born}(T)$ stems from the matrix element $M^{(1)}$
and survives in the Born approximation (hence the notation), whereas $I_\text{res}(T)$ corresponds
to the matrix element $M^{(2)}$ that is dominant for resonant impurities.
For $T>T_\text{BG}$ the term $I_0(T)$ scales linearly with temperature \cite{bistritzer2009,tse2009}:
\begin{equation}
I_0(T>T_\text{BG})= 4\pi \mathcal{W} T_\text{BG}^2 T,
\label{I0high}
\end{equation}
where
\begin{equation}
\label{calW}
\mathcal{W}=\frac{g^2\nu_F^2 k_B^3}{\hbar} \approx 0.05 \frac{n}{10^{12}\text{cm}^{-2}} \frac{W}{m^{2}K^{3}}
\end{equation}
and we have used $D=20$\,eV.

For the impurity-assisted terms in $I(T)$, we find:
\begin{align}
I_\text{Born}\left(T>T_\text{BG}\right)
&\simeq n_\text{imp}\mathcal{W}A_\text{Born}\cos^{2}\delta\sin^{2}\!\delta\ T^{3},
\label{IBorn} \\
I_\text{res}\left(T>T_\text{BG}\right)
&\simeq n_\text{imp}\mathcal{W}
 A_\text{res}\sin^{4}\!\delta\, \frac{T^5}{T_{BG}^2}
\ln^{2}\frac{s}{RT}.
\label{Ires}
\end{align}
Here, $A_\text{res}$ and $A_\text{Born}$ are the numerical coefficients:
\begin{equation}
A_\text{Born}=64\,\zeta(3)
\sum_{\alpha\beta}
\left\langle \left\vert \left\langle \mathbf{k},{\alpha}\right\vert
\left[
\hat{\Lambda},\left(  \hat{\bm{\sigma}}\hat{\mathbf{q}}\right)  \tau_{3}\right]
\left\vert \mathbf{k}^{\prime},{\beta}\right\rangle \right\vert^{2}\right\rangle_\text{FS}
\label{Aborn}
\end{equation}
and
\begin{equation}
A_\text{res}=\frac{3072\,\zeta(5)}{\pi^{2}}
\sum_{\alpha\beta}\left\langle \left\vert \left\langle \mathbf{k},{\alpha}\right\vert
\hat{\Lambda}\left\vert \mathbf{k}^{\prime},{\beta}\right\rangle \right\vert ^{2}\right\rangle_\text{FS}.
\label{Ares}
\end{equation}
For the short-range potential both the inter- and intra-valley transitions contribute
equally, whereas for the long-range potential only the inter-valley transitions are
allowed. As a result, the coefficients in these two models take different values:
\begin{equation}
A_\text{Born}=32\,\zeta(3),
\quad A_\text{res}
=\frac
{768\,\zeta(5)  }{\pi^{2}}\quad \text{ (short-range),} \label{short_res}
\end{equation}
\begin{equation}
A_\text{Born}=0,
\quad
A_\text{res}=\frac{6144\,\zeta(5)}{\pi^{2}}\quad
\text{(long-range),} \label{long_res}
\end{equation}
with $\zeta(x)$ the Riemann zeta-function.
As seen from Eqs.~(\ref{IBorn}) and (\ref{Ires}), for $T>T_{BG}$,
the resonant contribution to the energy flux has a $T^5$ temperature dependence,
which should be contrasted with the $T^3$ dependence of the Born term (the latter was calculated for weak impurities
in Ref.~\cite{song2012disorder}).

\begin{table*}[ht]
\begin{tabular}{|c||r|r|r|}
\hline
& \quad         $I_0$   \qquad\quad
& \quad         $I_\text{Born}$ \qquad\qquad\qquad\quad
& \quad         $I_\text{res}$   \qquad\qquad\qquad\qquad\quad            \\
\hline
\hline
\ $T\ll T_{\rm BG}$ \
&\quad $ \dfrac{T^4}{T_{\rm BG}} $ \quad\quad
&\qquad
$ \dfrac{n_{\rm imp}}{k_F^2}\dfrac{T^5}{T^2_{\rm BG}}\times \left\{\begin{array}{c}
                                     \delta^2, \quad \delta \ll 1  \\
                                     1, \quad  \delta \sim 1
                                   \end{array} \right.$
\qquad \quad
&\qquad                $\dfrac{n_{\rm imp}}{k_F^2} \dfrac{T^5}{T^2_{\rm BG}}\ln^2\!\left(\dfrac{1}{k_F R}\right)
           \times \left\{\begin{array}{c}
                                     \delta^4, \quad  \delta \ll 1  \\
                                     1, \quad  \delta \sim 1
                                   \end{array} \right.$
                                   \qquad \quad
        \\
\hline
\ $T\gg T_{\rm BG}$ \
&\qquad  $T\,T^2_{\rm BG} $   \quad\quad
&\qquad    $ \dfrac{n_{\rm imp}}{k_F^2}{T^3}
                    \times\left\{\begin{array}{c}
                                     \delta^2, \quad  \delta \ll 1  \\
                                     1, \quad  \delta \sim 1
                                   \end{array} \right.$
\qquad \quad
&\qquad          $\dfrac{n_{\rm imp}}{k_F^2} \dfrac{T^5}{T^2_{\rm BG}}\ln^2\!\left(\dfrac{T_{\rm BG}}{T k_F R}\right)
\times
\left\{      \begin{array}{c}
                                     \delta^4, \quad  \delta \ll 1  \\
                                     1, \quad  \delta \sim 1
                                   \end{array} \right.$
                                   \qquad
                                   \quad
          \\
          \hline
\end{tabular}
\caption{Scaling of different contributions to the heat flux $I(T)$ (in units of
$\mathcal{W}$, see Eq. (\ref{calW}) for low ($T\ll T_\text{BG}$)
and high ($T\gg T_{BG}$) temperatures. The term $I_0$ [Eqs.~(\ref{I0high}) and (\ref{I0low})] does not involve impurity scattering,
the terms $I_\text{Born}$ [Eqs.~(\ref{IBorn}) and (\ref{IBorn-low})] and $I_\text{res}$ [Eqs.~(\ref{Ires}) and (\ref{Ires-low})] describe
supercollisions. For each of the supercollision terms, results are given for weak ($\delta\ll 1$)
and strong ($\delta\sim 1$) scatterers.
\label{tab}}
\end{table*}

For the estimate of $I(T)$ at $T<T_{BG}$, we use matrix elements (\ref{M1-smallq}) and (\ref{M2-smallq}) for supercollisions,
which yields
\begin{eqnarray}
I_\text{Born}(T<T_\text{BG}) &\sim& \mathcal{W} \frac{n_{\rm imp}}{k_F^2}\, \sin^2\delta\cos^2\delta\, \frac{T^5}{T_{\rm BG}^2},\qquad
\label{IBorn-low}
\\
I_\text{res}(T<T_\text{BG}) &\sim& \mathcal{W} \frac{n_{\rm imp}}{k_F^2}\, \sin^4\delta \, \frac{T^5}{T^2_{\rm BG}} \ln^2\frac{1}{k_F R}.\qquad
\label{Ires-low}
\end{eqnarray}
We thus see that at low temperatures, both contributions to the impurity-assisted heat flux have a $T^5$ dependence.
For a strong impurity, $\tan \delta \gtrsim 1$, the second contribution to $I(T)$ always wins.
The term $I_0(T)$ at $T<T_{BG}$ behaves as \cite{chen2012}
\begin{equation}
I_0(T<T_\text{BG}) = \frac{8\pi^4}{15} \mathcal{W} \frac{T^4}{T_\text{BG}}.
\label{I0low}
\end{equation}
Thus, at low temperatures, $T<T_\text{BG}$, the term $I_0(T)$ scales as $T^4$, while both the supercollision terms
scale as $T^5$ (see Table~\ref{tab}).

In what follows, we assume that the system contains two types of impurities: weak ones
with the concentration $n_\text{imp}=n_{0}$ and phase shift $\delta_{0}\ll 1$ and a single resonant impurity
at position $r=0$.
In the resonant term, we substitute $n_\text{imp}
\rightarrow\delta\left(\mathbf{r}\right)$. As a result,
the function describing energy flux between electrons and phonons becomes $\mathbf{r}$-dependent,
\begin{equation}
I\left(\mathbf{r},T\right)=I_0(T)+I_\text{Born}\left(  T\right)  +
I_\text{res}\left(\mathbf{r},T\right).
\label{I_dec}
\end{equation}

We summarize the above results for the contributions to the heat flux between electrons and phonons at low ($T\ll T_\text{BG}$)
and high ($T\gg T_{BG}$) temperatures in Table~\ref{tab}.
These results will be used below for the analysis of the heat transfer  in
a weakly disordered graphene with a resonant impurity.

\section{Impurity-induced temperature distribution}
\label{s3}

\subsection{Heat-transfer equations in graphene}

We now turn to the effect of a single resonant impurity at $\mathbf{r}=0$ on the distribution of local
temperature in graphene, as measured in recent experiments \cite{2017arXiv171001486H}.
We assume electrons and phonons to be at the local thermodynamic equilibrium characterized by
temperatures $T_\text{e}$ and $T_\text{ph}$.
As we have shown in the previous Section, in the presence of a resonant impurity
on top of the background of weak impurities, there exist two contributions to the heat flux
between the electron and phonon systems: the homogeneous one, governed by weak disorder, and the local one, induced by the strong
scatterer. The electronic subsystem is electrically driven leading to the Joule heating. The overall heat balance
in the steady state is maintained by the coupling of the phonons to the thermal reservoir characterized by the base
(substrate) temperature $T_0$.

We assume for simplicity that the driving is weak, hence $T_\text{e}\approx T_\text{ph}\approx T$, and linearize all the non-linear
dependencies in the vicinity of $T$.
The spatial dependence of local temperatures in a macroscopic system
is governed by the following diffusion-type heat transfer equations:
\begin{align}
\left(C_\text{e}\partial_{t}-\kappa_\text{e}\nabla^{2}\right)  T_\text{e}
=&-\gamma\, \left(  T_\text{e}-T_\text{ph}\right)  - a\delta\left(  \mathbf{r}\right)  \left(  T_\text{e}-T_\text{ph}\right)
\nonumber
\\
&+E_{\infty}^{2}\sigma_0\left[  1+ b\delta\left(  \mathbf{r}\right)\right],
\label{eqTe}\\
\left( C_\text{ph}\partial_{t}-\kappa_\text{ph}\nabla^{2}\right)  T_\text{ph}
&=\gamma\, \left(  T_\text{e}-T_\text{ph}\right)
+a \delta\left(  \mathbf{r} \right) \left(  T_\text{e}-T_\text{ph}\right)
\nonumber
\\
&- \gamma_0\left( T_\text{ph}-T_{0}\right).
 \label{eqTp}
\end{align}
Here $C_\text{e,ph}$ are the heat capacities of electronic and phononic subsystems and
$\kappa_\text{e,ph}$ are the corresponding heat conductivities, and $\gamma_0$ quantifies the coupling to the bath.
Further, the parameters $\gamma$ and $a$ control the homogenous and the local (induced by the resonant impurity)
parts of the energy exchange between the electron and phonon systems, respectively.
If the homogeneous exchange is controlled by supercollisions assisted by weak impurities, the heat exchange rate $\gamma$
is given by
\begin{equation}
\gamma=\partial I_\text{Born}(T)/\partial T \,, \quad T>T_1,
\label{gamma}
\end{equation}
with $I_\text{Born}(T)$ given by Eq.~(\ref{IBorn}), and thus scales with temperature as $T^2$.
The temperature $T_1$ where this regime \cite{song2012disorder} is realized is given by
\begin{equation}
T_1=\sqrt{k_F l}\:T_\text{BG}.
\label{T1}
\end{equation}
For lower temperatures, the background electron-phonon scattering will be
determined by processes that do not involve impurities:
\begin{equation}
\gamma=\partial I_0(T)/\partial T, \quad T<T_1
\label{gamma_eph}
\end{equation}
 (see Table~\ref{tab}).
This will not make any change in the theory developed
in this Section, apart from a different scaling of $\gamma$.
The parameter $a$ is obtained from $I_\text{res}$ in Eq.~(\ref{I_dec}) in a similar way:
\be
a\delta\left(\mathbf{r}\right)=\partial I_\text{res}\left(\mathbf{r},T\right)/\partial T.
\label{a}
\end{equation}
where  we use Eq.~(\ref{Ires}) and Eq.~(\ref{Ires-low})  at $T>T_\text{BG}$  and $T<T_\text{BG},$ respectively [where $n_{\rm imp}$  is replaced with $\delta(\mathbf r)$].

The last term in Eq. (\ref{eqTe}) accounts for the Joule heat (with $\sigma_0$ being the conductivity outside the region of the strong scatterer
and $E_{\infty}$ the electric field at $r\to \infty$),
modified by the presence of resonant impurity. Here, we describe the effect of the strong scatterer by introducing phenomenologically a local term $b \delta\left(  \mathbf{r}\right)$. We will discuss microscopic origin and the characteristic magnitude of this term in connection with the physics of Landauer dipoles in Appendix \ref{App:A}.

Let us now analyze the stationary solutions of Eqs.~(\ref{eqTe}) and (\ref{eqTp})
perturbatively in the local heat-flux and Joule-heat terms induced by the resonant impurity. In the absence of the resonant
impurity ($a=b=0$), one obtains a homogeneous heating of the two subsystems:
\begin{equation}
T_\text{ph}^{\left(  0\right)}=T_{0}+\gamma_0^{-1}\sigma_0E_{\infty}^{2}
\label{Tph}
\end{equation}
and
\begin{equation}
T_\text{e}^{\left(  0\right)  }=T_{0}+\left(  \gamma^{-1}+\gamma_0^{-1}\right)
\sigma_0E_{\infty}^{2}.
\label{Te}
\end{equation}
It is worth noticing that at this level,
the phonon temperature $T_\text{ph}^{\left(  0\right)}$ is not sensitive to the rate of the electron-phonon heat exchange.

Next, we linearize Eqs.~(\ref{eqTe}) and (\ref{eqTp}) around the homogeneous solutions,
$T_\text{ph}\left(  r\right)=T_\text{ph}^{\left(0\right)}+\delta T_\text{ph}\left(  r\right)$
and $T_\text{e}\left(  r\right)=T_\text{e}^{\left(0\right)}+\delta T_\text{e}\left( r\right)$,
and find corrections to the phonon and electron temperatures induced by a single resonant impurity:
\begin{eqnarray}
\delta T_\text{ph}\left(  r\right)=T_{\ast}\left[ F_{1}(r)
+\frac{\gamma\eta}{\tilde{\gamma}}F_{2}(r)  \right]
\label{deltaTph}
\end{eqnarray}
and
\begin{equation}
\delta T_\text{e}\left(r\right)\!=\!T_{\ast}\!\!\left[\frac{\kappa_\text{ph}
\left(\eta-1\right)}{\kappa_\text{e}}F_{1}(r)
\!+\!\frac{(\gamma_{0}+\gamma)\eta-\gamma_{0}}{\tilde{\gamma}}F_{2}(r)  \right]\!.
\label{deltaTe}
\end{equation}
Here, the characteristic temperature scale $T_{\ast}$ is given by
\begin{equation}
T_{\ast}=\frac{\sigma_0E_{\infty}^{2}a}{2\pi\kappa_\text{ph}\gamma},
\label{Tstar}%
\end{equation}
the parameter
\begin{equation}
\eta=\frac{\gamma b}{a}
\label{eta}
\end{equation}
controls the relative importance of the local Joule heat at the scatterer,
and we have introduced
\begin{equation}
{\tilde{\gamma}}
=\gamma+(\gamma_0+\gamma)\frac{\kappa_\text{e}}{\kappa_\text{ph}}.
\label{Q}%
\end{equation}
To simplify the further analysis, we will assume below that the parameter $\eta$ is small, $|\eta| \ll 1$;
the validity of this assumption is supported by the microscopic analysis, see Appendix \ref{App:A}.
The spatial temperature distributions in Eqs.~(\ref{deltaTph}) and (\ref{deltaTe})  is governed by the functions
\begin{gather}
F_{1}\left(r\right)
=\frac{q_{1}^{2}\,K_{0}\left(q_{1}r\right)-q_{2}^{2}\,K_{0}\left(q_{2}r\right)}{q_{1}^{2}-q_{2}^{2}},
\label{F1}\\
F_{2}\left(r\right)
=(q_1^2+q_2^2)\frac{K_{0}\left(q_{2}r\right)-K_{0}\left(q_{1}r\right)  }{q_{1}^{2}-q_{2}^{2}}, \label{F2}%
\end{gather}
with $K_0(z)$ the modified Bessel function
and
\begin{equation}
q_{1,2}^{2}=\frac{\tilde{\gamma}}{2\kappa_\text{e}}
\mp\sqrt{\left(\frac{\tilde{\gamma}}{2\kappa_\text{e}}\right)^2-\frac{\gamma\gamma_0}{\kappa_\text{e}\kappa_\text{ph}}}.
\label{q12}%
\end{equation}

\begin{figure}[ptb]
\centering
\includegraphics[width=0.48\textwidth]{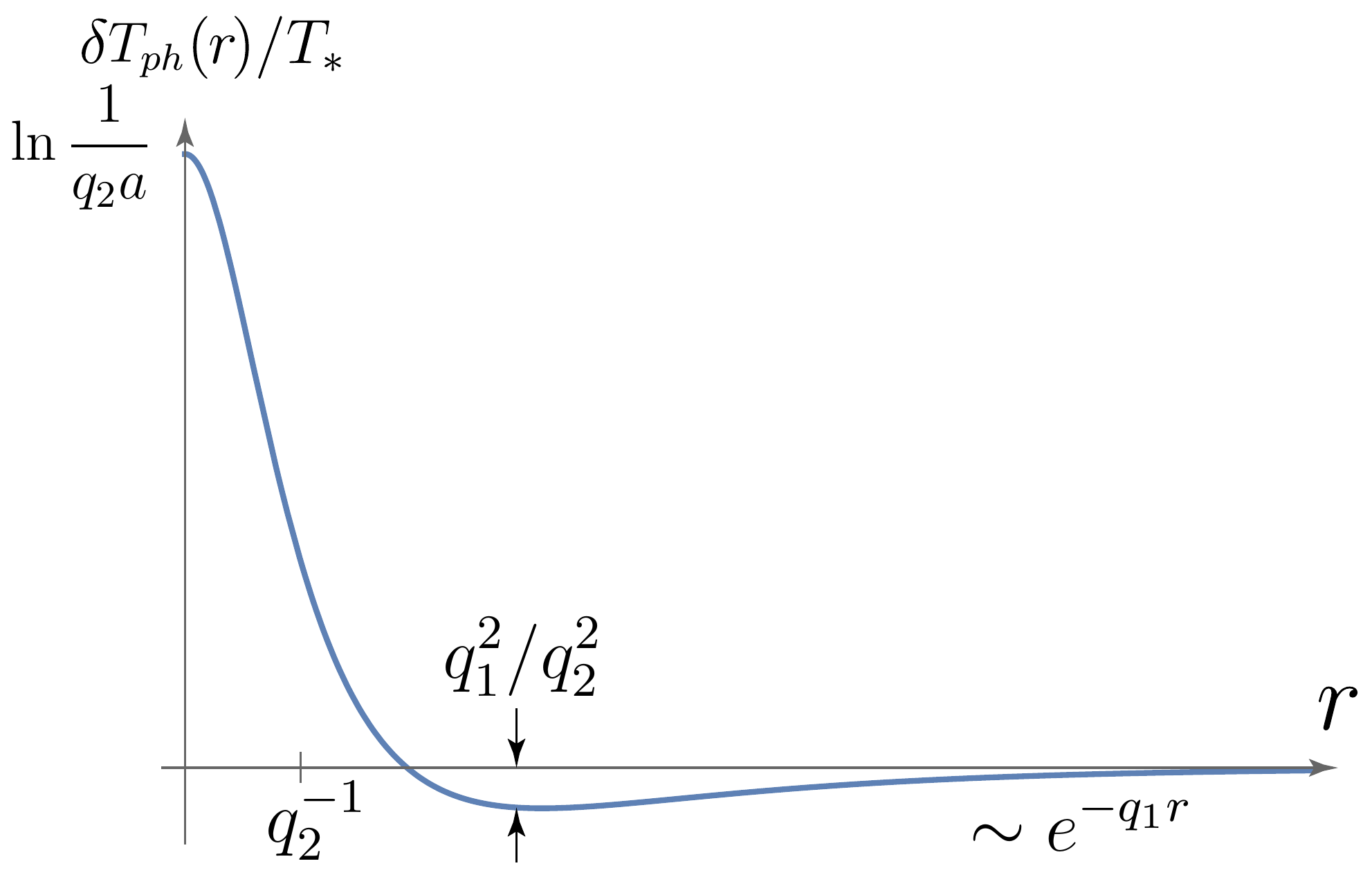}
\caption{Spatial dependence of the
phonon temperature $\delta T_\text{ph}(r)$ generated by supercollisions at a resonant impurity located at $r=0$.
}
\label{fig1}
\end{figure}\begin{figure}[ptb]
\centering
\includegraphics[width=0.48\textwidth]{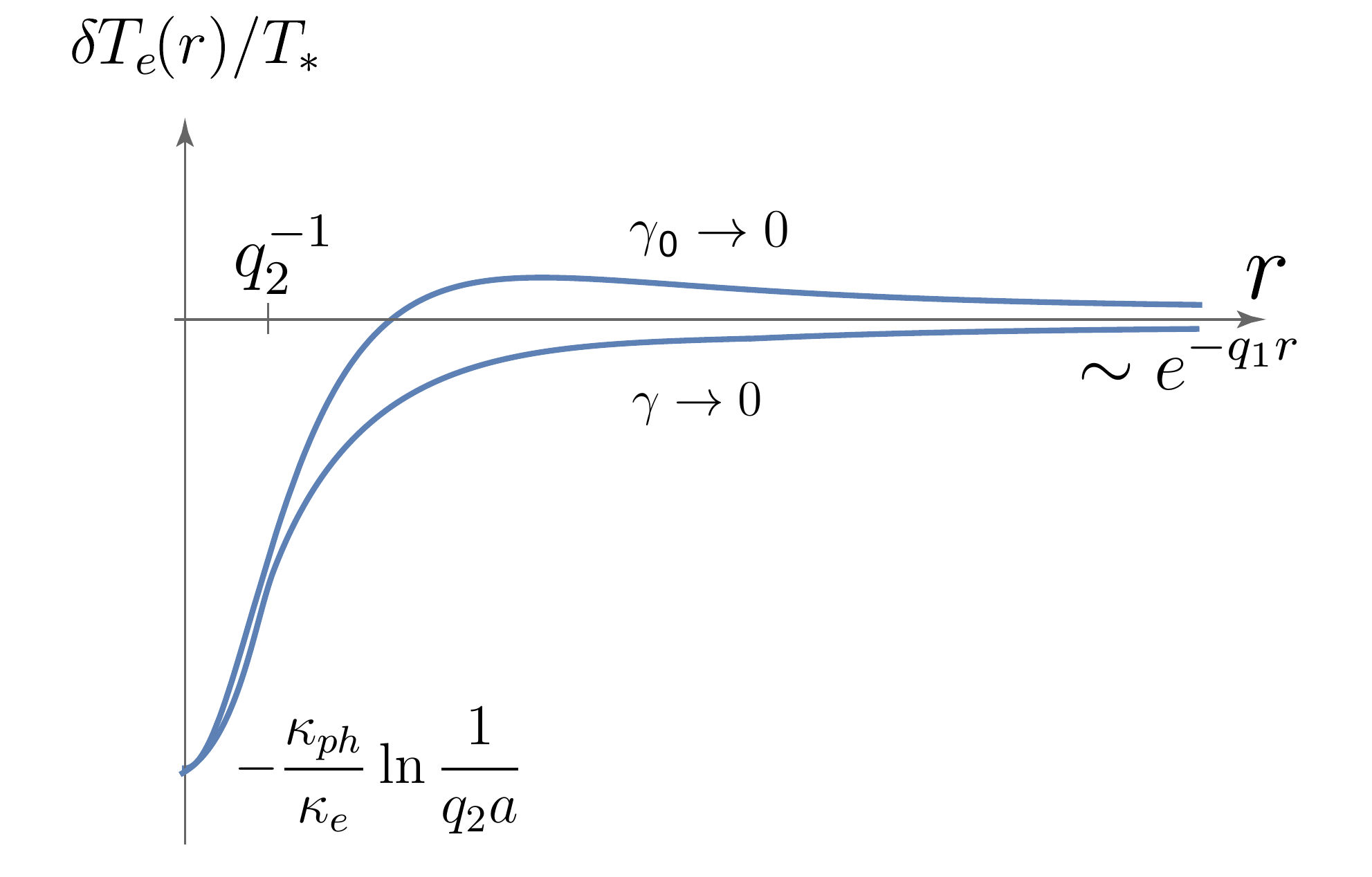}
\caption{Spatial dependence of the
electron temperature $\delta T_\text{e}(r)$ generated by supercollisions at a resonant impurity located at $r=0$.
}
\label{fig2}
\end{figure}

Equation~(\ref{q12}) defines the two spatial scales, $q^{-1}_{2}<q_{1}^{-1}$.
In the immediate vicinity of the impurity ($r\ll q_2^{-1}$), the functions
$F_1$ and $F_2$ produce a logarithmic singularity of
the local temperatures which is cut off by the ultraviolet scale $R$.
Further simplification is possible due to separation of scales in two limiting
cases. First, when the electron-phonon heat exchange is relatively weak in comparison to heat leakage to the substrate,
\begin{equation}
\gamma\ll\frac{\gamma_0}{1+\kappa_\text{ph}/\kappa_\text{e}},
\label{condition1}
\end{equation}
we have
\begin{equation}
q_{1}=\sqrt{\gamma/\kappa_\text{e}}, \quad q_{2}=\sqrt{\gamma_0/\kappa_\text{ph}} \,,
\label{qlimit1}
\end{equation}
with $q_{1}\ll q_{2}$.
In the opposite limit of a sufficiently strong homogeneous electron-phonon heat exchange,
\begin{equation}
\gamma\gg \frac{\gamma_0}{1+\kappa_\text{ph}/\kappa_\text{e}},
\label{condition2}
\end{equation}
we obtain
\begin{equation}
q_{1}=\sqrt{\frac{\gamma_0}{\kappa_\text{ph}+\kappa_\text{e}}},
\quad
q_{2}=\sqrt{\gamma\frac{  \kappa_\text{ph}+\kappa_\text{e}}{\kappa
_\text{e}\kappa_\text{ph}}  },
\label{qlimit2}
\end{equation}
and again $q_{1}\ll q_{2}$.

In any of these limits, we then find for the temperature profiles near the strong impurity:
\begin{eqnarray}
\delta T_\text{ph}\left(  r\ll q_{2}^{-1}\right)&\simeq&T_{\ast}\ln\frac{1}{q_2r},
\label{dTph0}\\
\delta T_\text{e}\left(  r\ll q_{2}^{-1}\right)
&\simeq&-T_{\ast}\frac{\kappa_\text{ph}}{\kappa_\text{e}}
\ln\frac{1}{q_2r}.
\label{dTe0}
\end{eqnarray}
Away from the impurity, the correction to the phonon temperature changes its sign:
\begin{equation}
\delta T_\text{ph}\left(  q_{2}^{-1}\ll r\ll q_{1}^{-1}\right) =-T_{\ast}
\frac{q_{1}^{2}}{q_{2}^{2}}K_{0}\left(  q_{1}r\right),
\end{equation}
as illustrated in Fig.~\ref{fig1}.

For the electron temperature (Fig.~\ref{fig2}), the sign of the correction away from the impurity
differs in the two limiting cases of small $\gamma$ and small $\gamma_0$:
\begin{eqnarray}
&& \delta T_\text{e}\left(q_{2}^{-1}\ll r\ll q_{1}^{-1}\right)  \nonumber \\
&& \qquad = T_{\ast} K_{0}\left(q_{1}r\right) \times
 \left\{
\begin{array}
[c]{c}%
\displaystyle-\frac{\kappa_{\rm ph}}{\kappa_\text{e}},
\quad \gamma\rightarrow0,\\[0.35cm]
\displaystyle\frac{\kappa_{\rm ph}}{\kappa_\text{e}+\kappa_{p}}\eta,\quad \gamma_{0}\rightarrow 0.
\end{array}
\right.
\label{delta-Te}
\end{eqnarray}

We have thus found that the presence of a single strong scatterer in a weakly disordered
graphene leads to the local heating (cooling) of the phonon (electron) subsystem in the vicinity of
the scatterer, mediated by the ``resonant supercollisions''. Away from the resonant
scatterer, the correction to the phonon temperature changes its sign.
The reason for this is essentially the energy conservation. Indeed, the resonant
supercollision leads to a local enhancement of release of the Joule heat accumulated
by the electron system. This should be compensated by some reduction of the energy released
to phonons further away from the scatterer.
Below, we will estimate the magnitude of the effect and discuss its experimental implications.

\subsection{Estimates for the characteristic temperature and length scales}
\label{s3B}

In this Section, we present estimates for the magnitude of the effect of resonant supercollision cooling
and for the characteristic spatial scales of the temperature distribution around a strong scatterer
in graphene. 

From Eq.~(\ref{Tstar}), assuming that $\gamma$ is dominated by processes without supercollisions
(clean samples, $T>T_1$)
and using expressions for the electron-phonon heat exchange from Table~\ref{tab},
we write for the characteristic magnitude of temperature variations:
\begin{equation}
T_{\ast}
\sim
\sin^{4}\!\delta\,
\frac{j^{2}_0}{\sigma_0\kappa_\text{ph} k_F^2}
\times \left\{\begin{array}{cc}
           \! \dfrac{T}{T_\text{BG}}\ln^2\dfrac{1}{k_F R},\ & T<T_\text{BG},\\[0.2cm]
           \!\! \left(\dfrac{T}{T_\text{BG}}\right)^4\ln^2\dfrac{T_\text{BG}}{k_FRT},\ & T>T_\text{BG},
           \end{array}\right.
\label{T_star_fin}
\end{equation}
where $j_0=\sigma_0 E_\infty$ is the current density far away from the scatterer.

Below we will assume that the impurity is near resonance and thus set $\sin \delta \sim 1$ for estimates.
A naive estimate for the parameter $b$ for a resonant impurity is $b \sim l /k_F$.
A simple way to obtain this estimate is to assume that the resistivity is determined by a finite concentration
of such resonant impurities and dividing the dissipated heat by the number of impurities.
If this estimate would be correct, we would have $\eta$ comparable to unity for temperatures around $T_{\rm BG}$.
It turns out, however, that this naive estimate is incorrect in the 2D case, as explained in Appendix \ref{App:A},
and $\eta$ is in fact much smaller. We thus discard it in our estimates below.

For estimates, we use for the parameters entering Eq.~(\ref{T_star_fin})
representative values suggested by the experiment \cite{2017arXiv171001486H}:
\begin{gather}
l\approx 1\,\mu\text{m},\quad n\approx 10^{12}\text{cm}^{-2}.
\end{gather}
With these values, we estimate
\begin{equation}
k_F=\sqrt{\pi n}\approx 1.8\cdot10^{8}\,\text{m}^{-1},
\quad
\sigma_0=\frac{e^2}{\pi\hbar}k_Fl \approx10^{-2}\,\Omega^{-1}.
\end{equation}
For the bias current, we use
$j_0\approx 1\,{\text{A}}/{\text{m}}$.
With the above value of carrier density, we have
\begin{equation}
T_\text{BG}\approx 27\,\text{K}, \quad T_1\approx 360\,\text{K},\quad  \mathcal{W}\approx 0.05\frac{\text{W}}{\text{m}^2 \text{K}^3}.
\end{equation}
Below we perform estimates for two values of temperature, corresponding to different regimes of temperature: $T=10$K ($T<T_\text{BG}$) and $T=50$K ($T>T_\text{BG}$).

Let us estimate the homogenous electron-phonon exchange rate $\gamma$
and the phonon-substrate~\cite{chen2009thermal,persson2010heat,Li17} $\gamma_0$ cooling rate.
At $T \sim 50\,\text{K}\ll T_1$, the homogeneous electron-phonon heat exchange is dominated by $I_0$ rather than by
supercollisions with weak impurities, see Eqs.~(\ref{T1}) and (\ref{gamma_eph}),
and can be estimated according to Eq.~(\ref{I0high}):
\begin{equation}
\gamma\approx 2.3\cdot 10^4\,\frac{\text{W}}{\text{m}^{2}\text{K}},\quad \gamma_0\approx 5\cdot 10^{7}\frac{\text{W}}{\text{m}^{2}\text{K}}.
\end{equation}
At $T=10$K, we have the low-temperature regime ($T<T_\text{BG}$), so that electron-phonon heat exchange is given by Eq.~(\ref{I0low}):
\begin{equation}
\gamma\approx 0.9\cdot 10^3 \frac{\text{W}}{\text{m}^{2}\text{K}},\quad \gamma_0\approx 5\cdot 10^{6}\frac{\text{W}}{\text{m}^{2}\text{K}}.
\end{equation}

Next, we estimate the homogenous overheating of phonons and electrons from the base temperature $T_0$
\begin{equation}
\delta T_{\textrm{el,ph}}^{(0)} = T_{\textrm{el,ph}}^{(0)}-T_0.
\end{equation}
The Joule heat is found to be $\sigma_0E_{\infty}^2=100\,\text{W}/\text{m}^2$.
This, together with above estimates for the cooling rates, gives
\begin{equation}
\delta T_{\textrm{el}}^{(0)}=5\,\mu \text{K},\quad \delta T_{\textrm{ph}}^{(0)}=2 \mu \text{K}
\end{equation}
for $T=50K$ and
\begin{equation}
\delta T_{\textrm{el}}^{(0)}=0.1\,\text{K},\quad \delta T_{\textrm{ph}}^{(0)}=20\, \mu \text{K}
\end{equation}
for $T=10K$.

For a quantitative estimate of the magnitude of the impurity-induced overheating of phonons
and respective spatial scales, we need to estimate the phonon and electron thermal conductivities.
For phonons, considering the graphene layer and the boron-nitride substrate (of 40 nm thickness) as a combined 2D system,
we use the results for boron nitride from Refs.~\cite{simpson1971} and \cite{Jo2013}.
The electronic heat conductivity can be estimated from the Wiedemann-Franz law,
$$\kappa_\text{e}=\frac{\pi^2 k_B^2}{3 e^2} \sigma_0T.$$
As a result, we get
\begin{equation}
\kappa_\text{ph}\approx 4\cdot10^{-7}\,\frac{\text{W}}{\text{K}},\quad \kappa_\text{e} \approx 1.2 \cdot10^{-8}\frac{\text{W}}{\text{K}}
\end{equation}
at $T=50$K and
\begin{equation}
\kappa_\text{ph}\approx 7\cdot10^{-9}\,\frac{\text{W}}{\text{K}},\quad \kappa_\text{e}\approx 2.4 \cdot10^{-9}\frac{\text{W}}{\text{K}}
\end{equation}
at $T=10$K.
The characteristic magnitude of the temperature variation induced by the scatterer can be quantified by the
following parameter:
\begin{equation}
\alpha_*=\frac{T_{\ast}}{\delta T_\text{ph}^{(0)}}=\frac{a}{2\pi\kappa_\text{ph}}\frac{\gamma_0}{\gamma}.
\end{equation}
Above, we have already estimated all relevant quantities apart from $a$, characterizing the impurity-assisted electron-phonon cooling rate.
It can be written as follows:
\be
a= \frac{5A_\text{res}}{k_F^2}\mathcal{W}\frac{T^4}{T_\text{BG}^2}\ln^2\frac{T_\text{BG}}{k_F R\, \text{max}\{T,T_\text{BG}\} }.
\ee
At $50$K it becomes (with $A_\text{res}\approx 600$ for long-range impurities):
\be
a\approx 4\cdot 10^{-10}\frac{\textrm{W}}{\textrm{K}}
\ee
and at $10$K:
\be
a\approx 10^{-12}\frac{\textrm{W}}{\textrm{K}}.
\ee
Combining all the above estimates, we find
\begin{eqnarray}
\alpha_*\approx \left\{\begin{array}{cc}
            0.4,\quad & T=50\,\text{K},\\
            0.2,\quad & T=10\,\text{K}.
           \end{array}\right.
\end{eqnarray}
The absolute values of the magnitude of the local temperature change read:
\begin{equation}
T_{\ast}(50\text{K})\approx 1\mu \text{K},\quad T_{\ast}(10\text{K})\approx 4\mu \text{K}.
\label{Tstar1050}
\end{equation}

We proceed now with the analysis of the characteristic spatial scales.
At $T=50$K we are in the regime (\ref{condition1}), in which the spatial scales are given by Eq.~(\ref{qlimit1}).
Combining the above values, we estimate the two spatial scales in the temperature profile at $50$K as
\begin{equation}
q_{1}^{-1}\approx {7\cdot 10^{-7}}\text{m},\quad q_{2}^{-1}\approx 10^{-7}\text{m}.
\label{q1-est}
\end{equation}
At $T=10$K, the condition (\ref{condition1}) is still fulfilled and we estimate
\begin{equation}
q_{1}^{-1}\approx {1.5\cdot 10^{-6}}\text{m},\quad q_{2}^{-1}\approx  4\cdot10^{-8}\text{m}.
\label{q1-est10}
\end{equation}
It is worth mentioning that our quasi-2D approximation for the $40$nm-thick slab of graphene and boron nitride
turns out to be at the border of applicability, since the slab thickness is comparable to the characteristic
size $q_{2}^{-1}$ of the temperature variation. This, in particular, implies that the actual value of $T_\ast$
may be a few times larger than that given by our estimates (\ref{Tstar1050}), while the size of the overheated region
as seen at the surface of the quasi-2D slab is expected to be somewhat larger than our 2D value of $q_{2}^{-1}$.

Finally, let us compare our results with experimental findings of Ref.~\cite{2017arXiv171001486H}.
First, the overall magnitude and sign of the effect are shown in Fig.~S5 of Ref.~\cite{2017arXiv171001486H},
with $\delta T>0$ and $\delta T\sim 5\,\mu$K, which is in rough agreement with our estimates, see Eq.~(\ref{Tstar1050}).
Next, the dependence of the excess temperature on the electrical current
(illustrated in Fig.~S9C of Ref.~\cite{2017arXiv171001486H}) is
quadratic, consistent with our Eq.~(\ref{T_star_fin}).
Finally, according to our Eqs.~(\ref{q1-est}), (\ref{q1-est10}) the size of the overheated region
$q_{2}^{-1}$ is about a few tens nanometers. The distances between the tip and the impurity at which an enhancement of temperature was detected in Ref.~\cite{2017arXiv171001486H} were of the order of or smaller than 100 nm, so that the measurement point was indeed
located in the ``overheated'' part in our Fig.~\ref{fig1}.

\section{Summary}
\label{s4}

In this work, we have studied the effect of strong (resonant) impurities on the heat transfer in a
coupled electron-phonon system in disordered graphene. Our key results can be summarised as follows.

First, we have investigated in detail how a strong impurity
modifies locally the electron-phonon heat exchange through the ``resonant-supercollision''
mechanism. The result is given by Eqs.~(\ref{Idef}) and (\ref{Ires})
and in Table~\ref{tab}.
For strong impurities, the contribution of supercollisions to the
function $I(T)$ describing the energy flow between electrons and phonon scales with
temperature as $T^5$, in contrast to the $T^3$ behavior found for weak impurities.

Second, we have explored the local modification of heat transfer induced by a resonant scatterer in a weakly disordered graphene
and calculated the spatial temperature profile around the scatterer under electrical driving.
The characteristic profiles of the phonon and electron temperature around the scatterer
are illustrated in Figs.~\ref{fig1} and \ref{fig2}.
The sign, magnitude, and characteristic spatial scale of the local temperature distribution of phonons are consistent
with the recent experimental findings on imaging resonant dissipation from individual atomic defects
reported in Ref.~\cite{2017arXiv171001486H}.

When we were preparing the manuscript for publication,  the preprint \cite{2017arXiv171001486H} appeared with
theoretical results that partly overlap with our analysis of ``resonant supercollisions".

\acknowledgments

We thank E. Zeldov for providing us with unpublished experimental results and for insightful discussions that stimulated this work.
We are also grateful to V. Khrapai and A. Finkel'stein for interesting discussions.
The work was supported by the joint grant of the Russian Science Foundation
(Grant No. 16-42-01035) and the Deutsche Forschungsgemeinschaft (Grant No. MI 658-9/1).

\appendix
\section{Effect of a scatterer on local Joule heat}
\label{App:A}

In this Appendix, we discuss the local effect of a scatterer on the Joule heat. For transparency, we first calculate the
distribution of the Joule heat in a model system, where
a spherical region with radius $R$  and  conductivity
$\sigma_{\rm in}$ is inserted at the origin of coordinate ($\mathbf r=0$)
into an infinite medium  with the conductivity $\sigma_{\rm out}$ to which
a homogeneous electric field $\mathbf E_\infty$  is applied in $x$ direction.
Then we extend the result to the case of an arbitrary scatterer inserted in a homogeneous medium.
We start with the discussion of a three-dimensional (3D) case and then generalize the results to the 2D case.

\subsection{3D case}

As stated above, we consider first a spherical region with radius $R$, center  $\mathbf r=0$, and  conductivity
$\sigma_{\rm in}$ inserted into an infinite medium  with the conductivity $\sigma_{\rm out}$. Away from the scatterer, the electric field is $\mathbf E_\infty$  and points in $x$ direction.
The distribution of electrical current $\mathbf{j}(\mathbf{r})$ obeys the condition
$$\text{div}\,\mathbf{j}=0.$$
With the local relation between the current density and the electric field
$\mathbf{j}(\mathbf{r})=\sigma \mathbf{E}(\mathbf{r}),$
this condition is equivalent to
$\text{div}\,\mathbf{E}=0$
both inside and outside the spherical region.
This implies that electric charges can appear only at the sphere surface.
We search for the  distribution of electric field outside the spherical region
as a sum of the field $E_\infty$ at $r\to \infty$ and the field of dipole emerged at the boundary $r=R$.
We also assume that the field is homogeneous for $r<R$.
The electrical potential is written as
\begin{equation}
\Phi(\mathbf{r})=\left\{ \begin{array}{cc}
\displaystyle  - E_{\rm in}r\cos\theta, &\quad{\rm for}\quad r<R,\\[0.2cm]
\displaystyle  - E_{\infty}r\cos\theta-\frac{d_3}{r^2}\cos\theta, &\quad{\rm for}\quad r>R,
           \end{array}\right.
\label{phi3D}
\end{equation}
where $\theta$ is the angle between the direction of $E_\infty$ and $\mathbf r$, and $d_3$ characterizes the strength of a 3D dipole.
The matching conditions for the potentials and currents at the boundary read:
\begin{gather}
E_{\rm in}R=\frac{d_3}{R^2}+E_{\infty}R,
\label{DipoleEqs3D}\\
\sigma_{\rm in}E_{\rm in}=\sigma_{\rm out}\left(  E_{\infty}-\frac{2d_3}{R^{3}}\right),
\nonumber
\end{gather}
yielding
\begin{eqnarray}
E_{\rm in}&=&E_{\infty}\frac{3\sigma_{\rm out}}{\sigma_{\rm in}+2\sigma_{\rm out}},
\\
d_3&=&E_{\infty}R^{3}
\frac{\sigma_{\rm out}-\sigma_{\rm in}}
{\sigma_{\rm in} +2\sigma_{\rm out}}.
\label{DipoleRes3D}%
\end{eqnarray}
Next, we calculate the Joule heat dissipated inside and outside the spherical region.
The heat dissipated in the region  $r<R$ is given by
\begin{equation}
 P_{\rm in}= \sigma_{\rm in} E_{\rm in }^2 \frac{4\pi R^3}{3}
 =\frac{12 \pi E_{\infty}^2R^{3}\sigma_{\rm out}^2\sigma_{\rm in}}{(\sigma_{\rm in}+2\sigma_{\rm out})^2}.
 \end{equation}
The heat dissipated outside the ball,
$$P_{\rm out}= \int \limits_{r>R}d^3r \sigma E^2,$$
contains a contribution from homogeneous external field $\propto E_\infty^2$, a dipole contribution $\propto d_3^2$,
and the cross term $\propto E_\infty d_3.$  The first term diverges at large $r$, so that we introduce a large finite volume $V\gg R^3$
of the whole system.
The cross-term cancels out after integration over angles.
Then, after integration of the dipole contribution,
we obtain
\begin{eqnarray}
 && P_{\rm out}= \sigma_{\rm out} E_{\infty}^2 \left (V - \frac{4\pi R^3}{3}\right)+\frac{4\pi \sigma_\infty d_3^2 }{3R^3}
 \\
 \nonumber
 && =\sigma_{\rm out} E_{\infty  }^2 \left (V -  \frac{4\pi R^3}{3}\right)
 +\frac{8 \pi E_{\infty}^2R^{3} \sigma_{\rm out} (\sigma_{\rm out}-\sigma_{\rm in})^2}
 {3(\sigma_{\rm in}+2\sigma_{\rm out})^2}.
 \end{eqnarray}
Now, we can find the total change of the dissipated power induced by the insertion of the
spherical region  with the conductivity $\sigma_{\rm in} \ne \sigma_{\rm out}$:
\begin{eqnarray}
\delta P&=&P_{\rm in} +P_{\rm out} - \sigma_{\rm out} E_{\infty  }^2 V
\\
\nonumber
&=& \frac
{4 \pi R^{3}}{3} \sigma_{\rm out} E_{\infty  }^2 \frac{\sigma_{\rm in}-\sigma_{\rm out}}
{\sigma_{\rm in}+2\sigma_{\rm out}}.
\end{eqnarray}

We thus see that the total correction to the Joule heat is proportional to the
product of the current density at infinity, $j_0=\sigma_{\rm out} E_{\infty}$,
and the ``Landauer dipole'' strength $d_3$:
\begin{equation}
\delta P = c_3 j_0 d_3,
\label{3D-correction}
\end{equation}
with
\be
c_3 = - \frac{4 \pi}{3}.
\label{c3}
\ee

When the local inhomogeneity is created by an individual impurity
(not characterized by the conductivity $\sigma_\text{in}$), it gives rise to the Landauer dipole of the magnitude
\cite{landauer1957spatial,chu88,zwerger1991exact,sorbello1998landauer}
\begin{equation}
d_3 = \frac{3 \pi \hbar j_0 s_\text{tr}}{4 e^2 k_F^2},
\end{equation}
where $s_\text{tr}$ is the transport scattering cross-section of the impurity.
The local variation of the Joule heat due to insertion of the scatterer can be still
expressed in terms of this dipole moment via Eq.~(\ref{3D-correction}).
A  transparent derivation of this result is given below in Sec.~\ref{app-subsec3}.

\subsection{2D case}

Let us now turn to the explicit calculation of the Joule heat in a 2D electronic system
with a disk of radius $R$ characterized by the conductivity $\sigma_\text{in}$ distinct from
the background conductivity. In a 2D case, the electron charge distribution $n(\mathbf{r})$
around the disk is no longer homogeneous and the electric potential is related to $n(\mathbf{r})$ by
\begin{equation}
\Phi(\mathbf{r})=\int d^2 r^\prime \frac{n(\mathbf{r}^\prime)}{|\mathbf{r}-\mathbf{r}^\prime|}.
\end{equation}
For the electrical current one has to take into account the ``diffusive contribution'' determined by the gradient
of the concentration:
\begin{eqnarray}
\mathbf{j}(\mathbf{r}) &=& \sigma(\mathbf{r}) \mathbf{E}(\mathbf{r})-D(\mathbf{r})\bm{\nabla} n(\mathbf{r}), \nonumber \\
& = & \sigma(\mathbf{r}) \mathbf{E}_{\rm ec}(\mathbf{r}),
\label{current-2D}
\end{eqnarray}
where $D(\mathbf{r})$ is the local diffusion coefficient. The second line of Eq.~(\ref{current-2D}) expresses the current in terms of the ``electrochemical field'' $\mathbf{E}_{\rm ec}(\mathbf{r}) = -e\bm{\nabla} \Phi_{\rm ec}(\mathbf{r})$, where $\Phi_{\rm ec}(\mathbf{r})$ is the electrochemical potential
(below, we drop the subscript ``ec'').
In a full analogy with the 3D case, the electrochemical potential has the form
\begin{equation}
\Phi(\mathbf{r})
=\left\{\begin{array}{cc}
             -E_{\rm in}r\cos\theta, &\quad{\rm for}\quad r<R,\\[0.2cm]
        \displaystyle     - E_{\infty}r\cos\theta-\frac{d_2}{r}\cos\theta, &\quad{\rm for}\quad r>R,
           \end{array}\right.
\label{phi2D}
\end{equation}
with the correction introduced by the inhomogeneous conductivity
having a form of a ``2D dipole'' characterized by $d_2$.
The matching conditions at the disk boundary read
\begin{gather}
E_{\rm in}R=\frac{d_2}{R}+E_{\infty}R, \label{DipoleEqs2D}\\
\sigma_{\rm in}E_{\rm in}=\sigma_{\rm out}\left(  E_{\infty}-\frac{d_2}{R^{2}}\right),
\nonumber
\end{gather}
yielding
\begin{eqnarray}
E_{\rm in}&=&E_{\infty}\frac{2\sigma_{\rm out}}{\sigma_{\rm in}+\sigma_{\rm out}},\\
d_2&=&E_{\infty}R^{2}
\frac
{\sigma_{\rm out}-\sigma_{\rm in}}
{\sigma_{\rm in}+\sigma_{\rm out}}.
\label{DipoleRes2D}%
\end{eqnarray}

The heat dissipated inside the disk,  $r<R$,   is given by
 \begin{equation}
 P_{\rm in}= \sigma_{\rm in} E_{\rm in }^2 {\pi R^2}=
 \frac
{4 \pi E_{\infty}^2R^{2}  \sigma_{\rm out}^2\sigma_{\rm in}   }{(\sigma_{\rm in}
+\sigma_{\rm out})^2}.
 \end{equation}
The heat dissipated  outside the disk reads
\begin{eqnarray}
P_{\rm out}&=& \int \limits_{r>R}d^2 r \sigma_{\rm out} E^2
=
\sigma_{\rm out} E_{\infty  }^2 \left (\mathcal{S}-\pi R^2 \right) +\frac{\pi \sigma_{\rm out} d_2^2 }{R^2}
   \nonumber
 \\
 &=&
 \sigma_{\rm out} E_{\infty  }^2 \left (\mathcal{S} -  \pi R^2 \right)+ \frac
{\pi R^2 \sigma_{\rm out} E_{\infty}^2\left(  \sigma_{\rm out}-\sigma_{\rm in} \right)^2  }{(\sigma_{\rm in}
+\sigma_{\rm out})^2},
\nonumber\\
\label{P3D}
 \end{eqnarray}
where $\mathcal{S}$ is total area of the system.
Remarkably, in contrast to the 3D case, the total change of the dissipated power induced by the insertion of the disk
equals zero:
\begin{equation}
\delta P=P_{\rm in} +P_{\rm out} - \sigma_{\rm out} E_{\infty  }^2  S  \equiv 0.
\end{equation}

For an individual scatterer, the strength of the 2D dipole was calculated in Refs.~\cite{chu88} and \cite{zwerger1991exact}:
\begin{equation}
d_2 =   \frac{2\hbar j_0 s_\text{tr}}{e^2 k_F}.
\end{equation}
Naively, one would expect, in analogy with Eq.~(\ref{3D-correction}),
\begin{equation}
\delta P = c_2 j_0 d_2.
\label{2D-correction}
\end{equation}
It turns out, however, that in the 2D case the numerical coefficient $c_2$ vanishes,
\be
c_2 = 0.
\label{c2}
\ee
A general reason for this result is given below.

\subsection{General analysis of the Joule heat}
\label{app-subsec3}

Below we present a more general derivation of a relation between the strength of the dipole and the local change $\delta P$ of the dissipated power. This will allow us to see that the difference between 3D and 2D cases that we have observed for a model of a macroscopic spherical obstacle is in fact of general character. To this end, we write the expression for total dissipated power as follows
\begin{eqnarray}
P&=&\int d^2r~ \mathbf E_{\rm ec}(\mathbf r) \mathbf j (\mathbf r) =- \int d^2\mathbf r~ \mathbf j  \bm{\nabla} \Phi(\mathbf{r})
  \nonumber
  \\
    &=&\int d^2 r \left[-{\rm div}\,(\Phi \mathbf{j})+\Phi\, {\rm div} \mathbf{j}\right].
 \end{eqnarray}
Since ${\rm div \mathbf j}=0$ in the stationary case, we find the total Joule heat as a surface integral
\begin{equation}
P=\sigma_{\rm out}\oint  dS\, \Phi(\mathbf{r})\, \mathbf n \cdot \bm{\nabla} \Phi(\mathbf{r}).
\label{P-asympt}
\end{equation}
Here, $\mathbf n$ is the normal vector to this surface  and we took into account that
$\mathbf j=-\sigma_{\rm  out} {\nabla} \Phi$ away from the scatterer.
We see that the Joule heat can be fully expressed in terms of the asymptotics of the electric potential at large $r$.
Let us assume  that the integration surface in Eq.~\eqref{P-asympt}  is spherical with the radius $R^\prime$ much larger than the size of the scatterer.
Using Eqs.~\eqref{phi3D}, \eqref{phi2D} and \eqref{P-asympt}, we get
\begin{equation}
P\!=\!\left\{  \begin{array}{cc}\!\!
          \displaystyle    \sigma_{\rm out} \frac{4\pi {R'}^3}{3} \left( E_\infty - \frac{2d_3}{{R'}^3}\right)\!
          \left(E_\infty + \frac{d_3}{{R'}^3}\right), &\text{  3D  case}  \\
             \\
             \displaystyle
            \!\!  \sigma_{\rm out} \pi {R'}^2 \left( E_\infty - \frac{d_2}{{R'}^2}\right)\!
            \left(E_\infty + \frac{d_2}{{R'}^2}\right), &\text{  2D  case}
           \end{array}
\right.
\end{equation}
Now we send $R'$ to infinity. The term, proportional to $E_\infty^2$ yields the Joule heat in the absence of the obstacle.
The term proportional to the square of the dipole tends to zero.
Hence, only the cross terms (those proportional to $E_\infty$ and to the dipole strength)
may give a correction $\delta P$ to the homogeneous Joule heat in the limit $R'\to \infty$.
For the 3D case, we  reproduce Eqs.~\eqref{3D-correction}, \eqref{c3}. For the 2D case,
the cross terms mutually cancel and we find  $\delta P = 0$,  in agreement with Eq.~\eqref{c2}.

Importantly, this derivation is quite general as it only uses the dipole form of the potential
at large distances as well as locality of the conductivity and the homogeneity of the system away
from the scatterer. One can expect that fluctuations in positions of impurities surrounding a
considered scatterer (including associated quantum interference effects) will produce a finite
$\delta P$ also in the 2D case. This effect should be, however, parametrically small in the case
of a good metallic system.


\begin{thebibliography}{10}

\bibitem{rokni1995joule}
M.~Rokni and Y~Levinson,
Joule heat in point contacts,
Phys. Rev. B \textbf{52}, 1882 (1995).

\bibitem{halbertal2016nanoscale}
D.~Halbertal, J.~Cuppens, M.~Ben Shalom, L.~Embon, N.~Shadmi, Y.~Anahory, H.R.~Naren,
J.~Sarkar, A.~Uri,	Y.~Ronen, Y.~Myasoedov,	L.S.~Levitov, E.~Joselevich, A.K.~Geim, and	E.~Zeldov,
Nanoscale thermal imaging of dissipation in quantum systems,
Nature \textbf{539}, 407 (2016).



\bibitem{song2012disorder}
J.C.W.~Song, M.Y.~Reizer, and L.S.~ Levitov,
Disorder-assisted electron-phonon scattering and cooling pathways in
  graphene, Phys. Rev. Lett. \textbf{109}, 106602 (2012).

\bibitem{song2015energy}
J.C.W.~Song and L.S.~ Levitov,
Energy flows in graphene: hot carrier dynamics and cooling,
J. Phys.: Cond. Matt \textbf{27}, 164201 (2015).


\bibitem{betz2012supercollision}
A.C.~Betz, S.H.~Jhang, E.~Pallecchi, R.~Feirrera, G.~F{\`e}ve,
J.-M.~Berroir, and B.~Pla{\c{c}}ais,
Supercollision cooling in undoped graphene,
Nature Phys. \textbf{9}, 109 (2013).

\bibitem{mckitterick2016electron}
C.B.~McKitterick, D.E.~Prober, and M.J.~Rooks,
Elec-tron-phonon cooling in large monolayer graphene devices,
Phys. Rev. B \textbf{93}, 075410 (2016).


\bibitem{fong2013measurement}
K.C.~Fong, E.E.~Wollman, H.~Ravi, W.~Chen, A.A.~Clerk,
  M.D.~Shaw, H.G.~Leduc, and K.C.~Schwab,
Measurement of the electronic thermal conductance channels and heat
  capacity of graphene at low temperature,
Phys. Rev. X \textbf{3}, 041008 (2013).


\bibitem{2017arXiv171001486H}
D.~Halbertal, M.~Ben~Shalom, A.~Uri, K.~Bagani, A.Y. Meltzer,
  I.~Marcus, Y.~Myasoedov, J.~Birkbeck, L.S.~Levitov, A.K.~Geim,
  and E.~Zeldov, Imaging resonant dissipation from individual atomic defects in
  graphene, arXiv:1710.01486.


\bibitem{ostrovsky2006electron}
P.M.~Ostrovsky, I.V.~Gornyi, and A.D.~Mirlin,
Electron transport in disordered graphene,
Phys. Rev. B \textbf{74}, 235443 (2006).


\bibitem{pereira2006disorder}
V.M.~Pereira, F.~Guinea, J.M.B.~Lopes Dos~Santos,
N.M.R.~Peres, and A.H.~Castro Neto,
Disorder induced localized states in graphene,
Phys. Rev. Lett. \textbf{96}, 036801 (2006).

\bibitem{basko2008resonant}
D.M.~Basko,
Resonant low-energy electron scattering on short-range impurities in
  graphene,
Phys. Rev. B \textbf{78}, 115432 (2008).

\bibitem{titov2010charge}
M.~Titov, P.M.~Ostrovsky, I.V.~Gornyi, A.~Schuessler, and A.D.~Mirlin,
Charge transport in graphene with resonant scatterers,
Phys. Rev. Lett. \textbf{104}, 076802 (2010).


\bibitem{landauer1957spatial}
R.~Landauer,
Spatial variation of currents and fields due to localized scatterers
  in metallic conduction,
IBM J. Res. Develop. \textbf{1}, 223 (1957).

\bibitem{landauer1988spatial}
R.~Landauer,
Spatial variation of currents and fields due to localized scatterers
  in metallic conduction,
IBM J. Res. Develop. \textbf{32}, 306 (1988).

\bibitem{chu88}
R.S.~Sorbello and C.S.~Chu,
Residual resistivity dipoles,
electromigration, and electronic conduction in
metallic microstructures,
IBM J. Res. Develop. \textbf{32}, 58 (1988).

\bibitem{zwerger1991exact}
W.~Zwerger, L.~B{\"o}nig, and K.~Sch{\"o}nhammer,
Exact scattering theory for the Landauer residual-resistivity dipole,
Phys. Rev. B \textbf{43}, 6434 (1991).

\bibitem{sorbello1998landauer}
R.S.~Sorbello,
Landauer fields in electron transport and electromigration,
Superlattices Microstruct. \textbf{23}, 711 (1998).



\bibitem{homoth2009electronic}
J.~Homoth, M.~Wenderoth, T.~Druga, L.~Winking, R.G.~Ulbrich, C.A.~Bobisch, B.~Weyers,
  A.~Bannani, E.~Zubkov, A.M.~Bernhart, M.R.~Kaspers, and R.~M\"oller,
Electronic transport on the nanoscale: Ballistic transmission and
Ohm's law,
Nano Lett. \textbf{9}, 1588 (2009).

\bibitem{willke2015spatial}
P.~Willke, T.~Druga, R.G.~Ulbrich, M.A.~Schneider, and
  M.~Wenderoth,
Spatial extent of a Landauer residual-resistivity dipole in graphene
  quantified by scanning tunnelling potentiometry,
Nature Commun. \textbf{6}, 6399 (2015).


\bibitem{bistritzer2009}
R.~Bistritzer and A.~H. MacDonald, Electronic cooling in graphene, Phys. Rev. Lett. \textbf{102}, 206410 (2009).

\bibitem{tse2009}
W.K.~Tse and S.~Das~Sarma, Energy relaxation of hot Dirac fermions in graphene, Phys. Rev. B \textbf{79}, 235406 (2009).

\bibitem{chen2012}
W.~Chen and A.A~Clerk,
Electron-phonon mediated heat flow in disordered graphene,
Phys. Rev. B \textbf{86}, 125443 (2012).

\bibitem{chen2009thermal}
Z.~Chen, W.~Jang, W.~Bao, C.N.~Lau, and C.~Dames,
Thermal contact resistance between graphene and silicon dioxide,
Appl. Phys. Lett. \textbf{95}, 161910 (2009).

\bibitem{persson2010heat}
B.N.J.~Persson and H.~Ueba,
Heat transfer between weakly coupled systems: Graphene on a-SiO$_2$,
Europhys. Lett. \textbf{91}, 56001 (2010).

\bibitem{Li17} X.~Li, Y.~Yan, L.~Dong, J.~Guo, A.~Aiyiti, X.~Xu, and B.~Li,
Thermal conduction across a boron nitride and SiO$_2$ interface,
J. Phys. D: Appl. Phys. \textbf{50}, 104002 (2017).


\bibitem{simpson1971}
A.~Simpson and A.D.~Stuckes, The thermal conductivity of highly oriented pyrolytic boron nitride,
J. Phys. C: Solid State Phys. \textbf{4}, 1710 (1971).

\bibitem{Jo2013} See, e.g., I.~Jo, M.T.~Pettes, J.~Kim, K.~Watanabe, T.~Taniguchi, Z.~Yao, and L.~Shi,
Thermal Conductivity and Phonon Transport in Suspended Few-Layer Hexagonal Boron Nitride,
Nano Lett. \textbf{13}, 550 (2013) and references therein.








\end{thebibliography}
\end{document}